\documentclass[12pt]{article}   
\usepackage{epsfig}

\newcommand{\mysection}{\setcounter{equation}{0}\section}

\def\beq{\begin{equation}}
\def\eeq{\end{equation}}
\def\beqa{\begin{eqnarray}}
\def\eeqa{\end{eqnarray}}

\newlength{\dinwidth} \newlength{\dinmargin}
\begin{document}

\begin{center}
{\Large \bf Next-to-next-to-leading-logarithm resummation for $s$-channel 
single top quark production}
\end{center}
\vspace{2mm}
\begin{center}
{\large Nikolaos Kidonakis}\\
\vspace{2mm}
{\it Kennesaw State University, Physics \#1202,\\
1000 Chastain Rd., Kennesaw, GA 30144-5591}\\
\end{center}

\begin{abstract}
I present the next-to-next-to-leading-logarithm (NNLL) resummation of 
soft and collinear gluon corrections to single  
top quark production in the $s$ channel.
Attaining NNLL accuracy involves the calculation of the two-loop soft anomalous dimension for the partonic subprocesses. 
Finite-order expansions of the resummed cross section are calculated 
through next-to-next-to-leading order (NNLO).
Numerical results are presented for $s$-channel single top quark production 
at the Tevatron and the LHC, 
including the dependence of the cross sections on 
the top quark mass and the uncertainties in the theoretical prediction. 
The higher-order corrections are significant for energies 
at both colliders and they decrease the theoretical uncertainty. 
\end{abstract}

\mysection{Introduction}

The recent observation of single top quark production at the Tevatron 
\cite{D0st,CDFst,singletop} and 
the new era that the Large Hadron Collider (LHC) is ready to embark on 
have made accurate theoretical calculations of single top quark cross sections 
imperative.
The study of single top quark processes provides unique opportunities 
for understanding the 
electroweak properties of the top quark, including a direct measurement of the 
$V_{tb}$ CKM matrix element, and for further insights into electroweak
theory and future discoveries of new physics (for top physics reviews 
see Ref. \cite{topreview}).

The production of single top quarks can proceed via three distinct 
partonic processes that involve the exchange of a space-like $W$ boson 
($t$ channel), the exchange of a time-like $W$ boson ($s$ channel), 
and $W$ emission in association with a top quark ($tW$ channel). 
In this paper we concentrate on the $s$ channel.
In the $s$ channel we have lowest-order processes of the form 
$q{\bar q}' \rightarrow {\bar b} t$ (Fig. 1), which include 
the dominant process $u {\bar d} \rightarrow {\bar b} t$   
as well as processes involving the charm quark and Cabibbo-supressed 
contributions. 
The QCD corrections for $s$-channel production at next-to-leading order (NLO) 
are known at the differential level \cite{bwhl} and 
are found to increase the cross section
and stabilize the dependence on the factorization scale \cite{bwhl,HCSY}.

\begin{figure}
\begin{center}
\includegraphics[width=8cm]{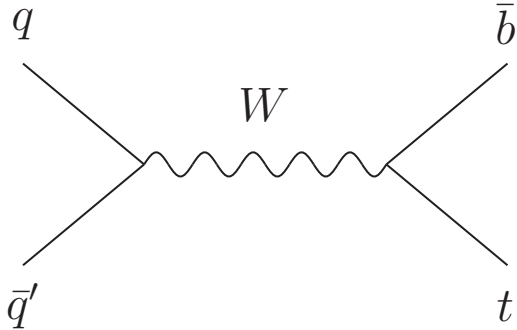}
\caption{\label{slo} Leading-order $s$-channel diagram for 
single top quark production.}
\end{center}
\end{figure}

Further improvement of the theoretical calculations was achieved in 
Ref. \cite{NKst} where the soft-gluon logarithms were resummed for 
single-top quark production processes at next-to-leading-logarithm (NLL) 
accuracy. NLL resummation requires the calculation of one-loop diagrams  
in the eikonal approximation. The higher-order soft-gluon contributions 
further increase the cross section at Tevatron and LHC energies 
\cite{NKst,NKstlhc,NKtop}.
Recent developments in the calculation of two-loop soft anomalous dimensions 
with massive and massless quarks \cite{NK2l,DPF092l} now allow the 
calculation of two-loop eikonal corrections and 
thus of next-to-next-to-leading-logarithm (NNLL) resummation for single 
top quark production.

In the next section we employ the resummation formalism of \cite{NKst} and 
extend it to NNLL accuracy. To achieve NNLL accuracy 
we calculate the soft anomalous dimension for $s$-channel single-top 
production through two-loops. 
We then expand the NNLL resummed cross section through 
next-to-next-to-leading order (NNLO) in the strong coupling, $\alpha_s$. 
NNLL resummation allows the determination of all 
soft-gluon terms at NNLO, thus improving the results of \cite{NKst} 
where only the first two powers of logarithms were fully computed. 
The approximate NNLO expression thus derived here is then used in the 
following sections to compute numerical results for the single top 
and single antitop cross sections at the Tevatron and the LHC.
 
\mysection{Threshold resummation}

In this section we present the analytical form of the 
resummed cross section for single top quark production in the $s$ channel.  
Details of the general resummation formalism for hard-scattering cross 
sections \cite{KOS,NNNLO} and the specific implementation for single top 
quark processes \cite{NKst,NKstlhc,NKtop} have been presented elsewhere, 
so here we explicitly show only the 
expressions directly relevant to NNLL single top quark 
$s$-channel production, without a detailed review. 

For the process $q+{\bar q}' \rightarrow {\bar b}+t$,
the partonic kinematical invariants are 
$s=(p_q+p_{{\bar q}'})^2$, $t=(p_q-p_{\bar b})^2$, $u=(p_{{\bar q}'}
-p_{\bar b})^2$, $s_4=s+t+u-m_t^2$, with $m_t$ the top quark mass while
the $b$-quark is taken to be massless \cite{NKst}. 
As we approach kinematical threshold the 
invariant $s_4$ approaches zero. The soft-gluon logarithms 
that appear in the perturbative partonic cross section are of the form 
$\ln^k(s_4/m_t^2)/s_4$. 
Resummation of the soft-gluon contributions 
is performed in moment space, where we define moments 
of the cross section by ${\hat\sigma}(N)=\int (ds_4/s) \;  e^{-N s_4/s} 
{\hat\sigma}(s_4)$,
with $N$ the moment variable. In the cross section the logarithms of $s_4$ 
transform into logarithms of $N$, which exponentiate.
The resummed cross section in moment space is derived by 
factorizing the cross section into hard, soft, and 
jet functions and solving their renormalization group equations
\cite{KOS}.
For $s$-channel single top production the resummed 
partonic cross section is then given by 
\beqa
{\hat{\sigma}}^{res}(N) &=&   
\exp\left[ \sum_{i=1,2} E(N_i)\right] \; 
\exp\left[ {E'}(N')\right] \; 
\exp \left[\sum_{i=1,2} 2 \int_{\mu_F}^{\sqrt{s}} \frac{d\mu}{\mu}\;
\gamma_{q/q}\left({\tilde N}_i, \alpha_s(\mu)\right)\right] \;
\nonumber\\ && \hspace{-10mm} \times \,
{\rm Tr} \left \{H^{q {\bar q}'\rightarrow {\bar b} t}
\left(\alpha_s(\sqrt{s})\right) \;
\exp \left[\int_{\sqrt{s}}^{{\sqrt{s}}/{\tilde N'}} 
\frac{d\mu}{\mu} \;\Gamma_S^{\dagger \, q {\bar q}'\rightarrow {\bar b} t}
\left(\alpha_s(\mu)\right)\right] \right.
\nonumber\\ && \quad \left.  \times \,
S^{q {\bar q}'\rightarrow {\bar b} t}\left(\alpha_s(\sqrt{s}/{\tilde N'})
\right) \; \exp \left[\int_{\sqrt{s}}^{{\sqrt{s}}/{\tilde N'}} 
\frac{d\mu}{\mu}\; \Gamma_S^{q {\bar q}'\rightarrow {\bar b} t}
\left(\alpha_s(\mu)\right)\right] \right\} \, .
\label{resHS}
\eeqa

The first exponent \cite{GS87,CT89} in the above expression 
resums soft and collinear corrections from the incoming quark and antiquark
\beq
E(N_i)=
\int^1_0 dz \frac{z^{N_i-1}-1}{1-z}\;
\left \{\int_1^{(1-z)^2} \frac{d\lambda}{\lambda}
A\left(\alpha_s(\lambda s)\right)
+D\left[\alpha_s((1-z)^2 s)\right]\right\} \, .
\label{Eexp}
\eeq
Here $N_1=N[(m_t^2-u)/m_t^2]$ and 
$N_2=N[(m_t^2-t)/m_t^2]$. 
The quantity $A$ has a perturbative expansion, 
$A=\sum_n (\alpha_s/\pi)^n A^{(n)}$. 
Here 
$A^{(1)}=C_F$ with $C_F=(N_c^2-1)/(2N_c)$ where $N_c=3$ is 
the number of colors,
while $A^{(2)}=C_F K/2$ with 
$K= C_A\; ( 67/18-\pi^2/6 ) - 5n_f/9$ \cite{KT82},
where $C_A=N_c$, and $n_f=5$ is the number of light quark flavors.            

Also $D=\sum_n (\alpha_s/\pi)^n D^{(n)}$, 
with $D^{(1)}=0$ in Feynman gauge ($D^{(1)}=-C_F$ in axial gauge) 
and \cite{CLS97}
\beq
D^{(2)}=C_F C_A \left(-\frac{101}{54}+\frac{11}{6} \zeta_2
+\frac{7}{4}\zeta_3\right)
+C_F n_f \left(\frac{7}{27}-\frac{\zeta_2}{3}\right) 
\eeq
in Feynman gauge where $\zeta_2=\pi^2/6$ and $\zeta_3=1.2020569\cdots$.

The second exponent \cite{GS87,CT89} resums soft and collinear corrections 
from the outgoing $b$-quark and can be written in the form \cite{NKVDD}
\beq
{E'}(N')=
\int^1_0 dz \frac{z^{N'-1}-1}{1-z}\;
\left \{\int^{1-z}_{(1-z)^2} \frac{d\lambda}{\lambda}
A \left(\alpha_s\left(\lambda s\right)\right)
+B\left[\alpha_s((1-z)s)\right]
+D\left[\alpha_s((1-z)^2 s)\right]\right\} \, ,
\label{E'exp}
\eeq
where  $N'=N (s/m_t^2)$ and $A$ and $D$ are defined above. 
Here $B=\sum_n (\alpha_s/\pi)^n B^{(n)}$ 
with $B^{(1)}=-3C_F/4$ and
\beq
B^{(2)}=C_F^2\left(-\frac{3}{32}+\frac{3}{4}\zeta_2-\frac{3}{2}\zeta_3\right)
+C_F C_A \left(\frac{77}{864}-\frac{11}{4}\zeta_2-\zeta_3\right)
+n_f C_F \left(\frac{23}{432}+\frac{\zeta_2}{2}\right).
\eeq

In the third exponent $\gamma_{q/q}$ is the moment-space 
anomalous dimension of the ${\overline {\rm MS}}$ 
parton density $\phi_{q/q}$ and it controls the factorization scale, $\mu_F$, 
dependence of the cross section. 
We have $\gamma_{q/q}=-A \ln {\tilde N}_i +\gamma_q$ where $A$ was 
defined above, ${\tilde N}_i=N_i e^{\gamma_E}$ 
with $\gamma_E$ the Euler constant, and the parton anomalous dimension
$\gamma_q=\sum_n (\alpha_s/\pi)^n \gamma_q^{(n)}$
where  $\gamma_q^{(1)}=3C_F/4$.

$H^{q {\bar q}'\rightarrow {\bar b} t}$ is the hard-scattering function 
while $S^{q {\bar q}'\rightarrow {\bar b} t}$ is the 
soft function describing noncollinear soft gluon emission \cite{KOS}.
The evolution of the soft function is controlled by the soft anomalous 
dimension $\Gamma_S^{q {\bar q}'\rightarrow {\bar b} t}$.
Here ${\tilde N'}={\tilde N}(s/m_t^2)$ with ${\tilde N}=N e^{\gamma_E}$.
Note that $H$, $S$, and $\Gamma_S$  are matrices in a basis 
consisting of color exchange (i.e. for the process with color indices 
$a+b \rightarrow c+d$ a color basis is $e_1=\delta_{ab}\delta_{cd}$ and 
$e_2=T^e_{ba} T^e_{dc}$) and 
the color trace is taken of their product, which   
at lowest order is the Born cross section.
We expand the soft anomalous dimension as
$\Gamma_S=\sum_n (\alpha_s/\pi)^n \Gamma_S^{(n)}$. 
Because of the simple color structure of the hard scattering
for single top $s$-channel production, 
the hard and soft matrices take a very simple form and only the first 
diagonal element of the one-loop soft anomalous dimension matrix, 
$\Gamma_{S\, 11}^{(1)}$, is needed in the NNLO expansion at NLL accuracy. 

\begin{figure}
\begin{center}
\includegraphics[width=13cm]{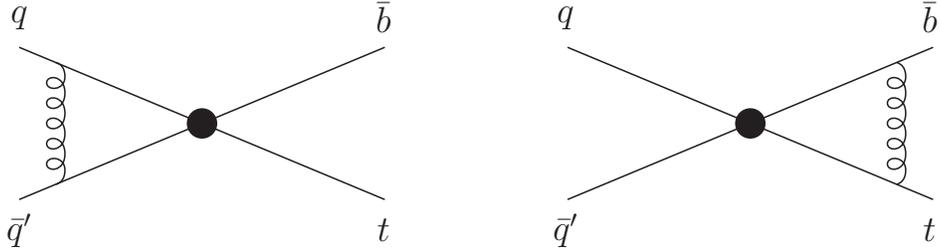}
\caption{\label{s1l} One-loop vertex-correction eikonal diagrams 
for $\Gamma_{S \, 11}^{(1)}$.} 
\end{center}
\end{figure}

The required element $\Gamma_{S\, 11}^{(1)}$ of the one-loop soft 
anomalous dimension for $s$-channel single-top production, necessary 
for NLL accuracy, was calculated 
in Ref. \cite{NKst}. The calculation involves one-loop eikonal diagrams, 
which include the one-loop vertex corrections in Fig. 2 plus the one-loop  
self-energy correction for the top quark line in Fig. 4 (top left diagram).
We employ dimensional regularization and determine the soft anomalous dimension
from the coefficients of the ultraviolet poles.
The result presented in \cite{NKst} used the axial gauge for the gluon 
propagator. Here we use  
the Feynman gauge and thus the result takes the slightly different form
\beq
\Gamma_{S\, 11}^{(1)}=C_F \left[\ln\left(\frac{s-m_t^2}{m_t\sqrt{s}}\right)
-\frac{1}{2}\right] \, .
\eeq
The change is of course compensated by different expressions for the 
$D$ coefficient (see above) so that the final result for the resummed 
cross section is identical in the two gauges.

The off-diagonal one-loop elements are needed in the NNLO expansion 
at NNLL accuracy.
We find
\beq
\Gamma_{S\, 21}^{(1)}=\ln\left(\frac{u(m_t^2-u)}{t(m_t^2-t)}\right)\; ,
\quad \quad
\Gamma_{S\, 12}^{(1)}=\frac{C_F}{2N_c} \, \Gamma_{S\,21}^{(1)} \, .
\eeq

\begin{figure}
\begin{center}
\includegraphics[width=9cm]{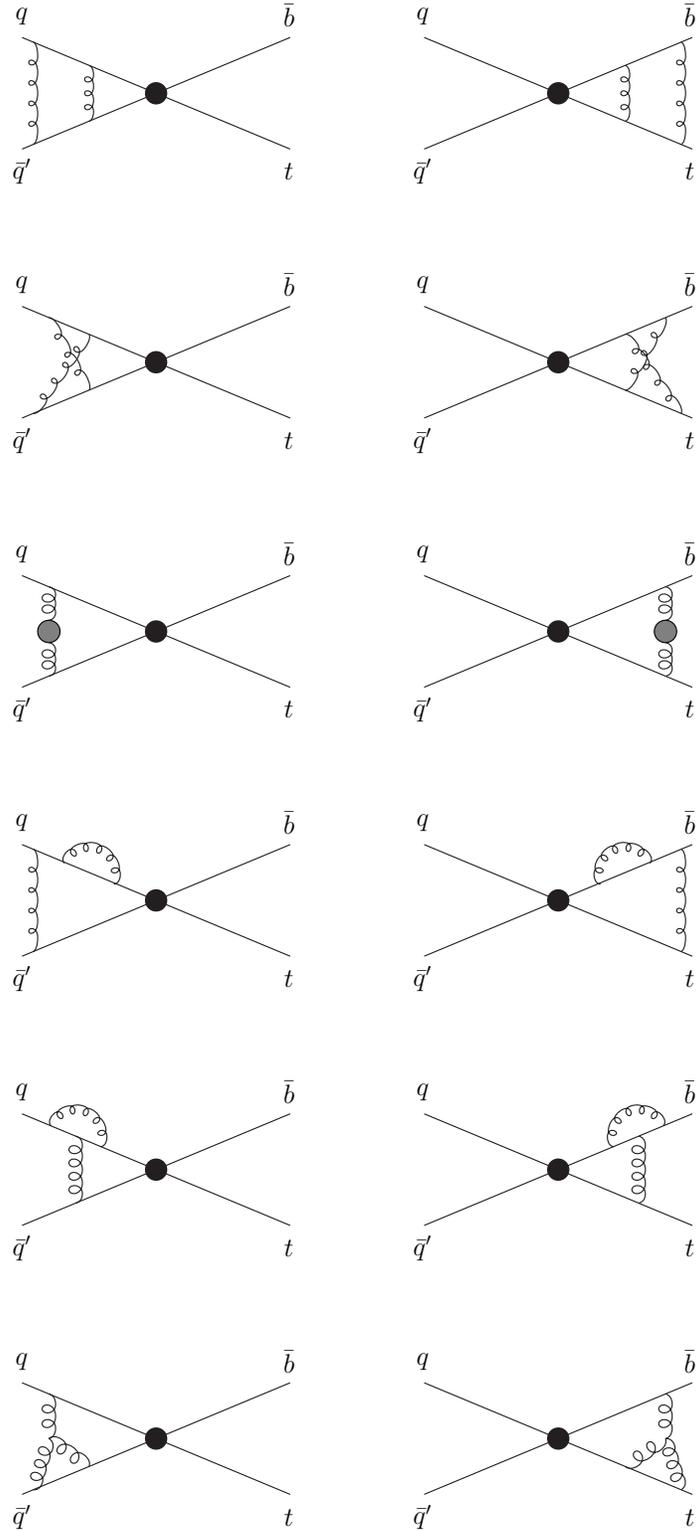}
\caption{\label{s2l} Two-loop vertex-correction eikonal diagrams  
for $\Gamma_{S \, 11}^{(2)}$. The shaded blob in the fifth and sixth diagrams 
denotes quark, gluon, and ghost loops.
Note that for each of the bottom six diagrams 
there is an additional diagram (not shown) with the gluon attached to the 
opposite quark line.} 
\end{center}
\end{figure}

At NNLL accuracy we also need to calculate the two-loop soft anomalous 
dimension. In the NNLO expansion at NNLL accuracy we need the  
element $\Gamma_{S \; 11}^{(2)}$ which we calculate by 
evaluating two-loop eikonal diagrams 
involving the four quarks in the hard scattering. Since only one of the 
eikonal lines (the top quark line) has mass, we can use the results of 
Ref. \cite{NK2l,DPF092l}, which involve 
pairs of massive quarks, and take the massless limit for one 
or two quarks \cite{DPF092l}. 
There are many eikonal diagrams to be calculated at two loops: 
the two-loop vertex correction diagrams shown in Fig. 3 plus the two-loop  
self-energy corrections for the top quark line in Fig. 4.
Again, we employ dimensional regularization and calculate the soft 
anomalous dimension from the coefficients of the ultraviolet poles 
of the two-loop diagrams.
Note that diagrams for this process involving three eikonal lines do 
not contribute. 
This is because three-parton diagrams with at least two massless eikonal lines
vanish \cite{ADS}. 
Analyzing all the diagrams we find 
\beq
\Gamma_{S\, 11}^{(2)}=\frac{K}{2}\Gamma_{S\, 11}^{(1)}
+C_F C_A \frac{(1-\zeta_3)}{4}
\eeq
where $K$ is the two-loop constant defined previously. 
The two-loop result above is written in terms of the one-loop element 
$\Gamma_{S\, 11}^{(1)}$. 

\begin{figure}
\begin{center}
\includegraphics[width=9.5cm]{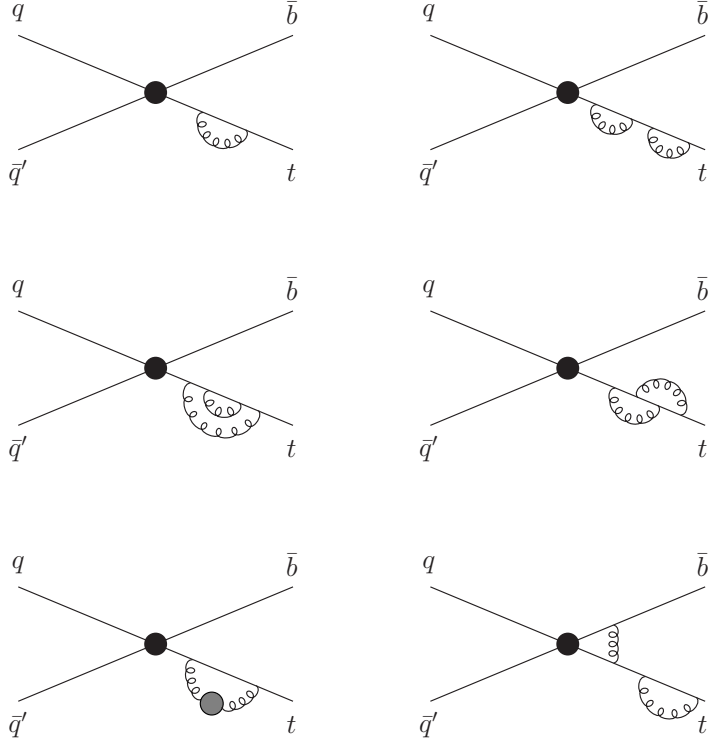}
\caption{\label{s2ls} One-loop (top left diagram) and two-loop top-quark 
self-energy eikonal diagrams. The shaded blob in the bottom left diagram 
denotes quark, gluon, and ghost loops.} 
\end{center}
\end{figure}

The resummed cross section, Eq. (\ref{resHS}),  
can be expanded in the strong coupling, $\alpha_s$, 
and inverted to momentum space, thus providing fixed-order results 
for the soft-gluon corrections.
The NLO expansion of the resummed cross section after inversion to 
momentum space is 
\beq
{\hat{\sigma}}^{(1)} = \sigma^B \frac{\alpha_s(\mu_R)}{\pi}
\left\{c_3\, {\cal D}_1(s_4) + c_2\,  {\cal D}_0(s_4) \right\} \, ,
\label{NLOmaster}
\eeq
where $\mu_R$ is the renormalization scale and we use the notation
${\cal D}_k(s_4)=[\ln^k(s_4/m_t^2)/s_4]_+$ for the plus distributions 
involving logarithms of $s_4$.
Here $\sigma^B$ is the Born term, and the coefficient of the leading term 
is
\beq
c_3=3 \, A^{(1)}\, .
\label{c3}
\eeq 
The coefficient of the next-to-leading term, $c_2$, can be written as
 $c_2=c_2^{\mu}+T_2$, 
with
\beq
c_2^{\mu}=-2 A^{(1)} \ln\left(\frac{\mu_F^2}{m_t^2}\right)
\eeq
denoting the terms involving logarithms of the scale, and  
\beq
T_2=-2 \, A^{(1)} \, \ln\left(\frac{(m_t^2-t)(m_t^2-u)}{m_t^4}\right)+3 D^{(1)}
-3 A^{(1)} \ln\left(\frac{m_t^2}{s}\right)
+B^{(1)} +2 \Gamma_{S\, 11}^{(1)}\, 
\label{c2n}
\eeq
denoting the scale-independent terms.
As discussed in \cite{NKst} the expansion can also determine 
the terms involving logarithms of the factorization scale in the coefficient, 
$c_1$, of the $\delta(s_4)$ terms. If we denote these terms as
$c_1^{\mu}$, then
\beq
c_1^{\mu}=\left[A^{(1)}\, \ln\left(\frac{(m_t^2-t)(m_t^2-u)}{m_t^4}\right) 
-2 \gamma_q^{(1)}\right]\ln\left(\frac{\mu_F^2}{m_t^2}\right) \, .
\label{c1mu}
\eeq
The full virtual terms are not derivable from resummation, which addresses 
soft-gluon contributions, but can be taken from the complete NLO calculation.

As has been shown in \cite{NKst,NKstlhc} the NLO expansion of the
resummed cross section approximates well the complete NLO result 
for both Tevatron and LHC energies. In fact when damping factors are used to 
limit the soft-gluon contributions far away from threshold, as was also used 
for $t{\bar t}$ production \cite{NKRV}, then the approximation is excellent. 
Thus, it is clear that for $s$-channel single top quark production the 
soft-gluon corrections dominate the cross section while contributions from other classes of 
corrections are negligible, 
so higher-order expansions of the soft-gluon resummed cross section can be reasonably expected to 
closely approximate the complete cross section.
This is an important consideration since it is not always true for every process 
that the soft corrections dominate the cross section. For example, for single top production 
via the $t$ channel at the LHC it was shown in \cite{NKstlhc} that this is not so. 
In such cases other classes of corrections, such as hard-gluon and virtual terms, can be important. 
In Higgs production via $b{\bar b} \rightarrow H$ (and $gg\rightarrow H$),
for example, it was shown that purely collinear corrections are large and that together 
with the soft corrections they 
provide an excellent approximation to the complete corrections at both NLO and NNLO \cite{NKHiggs}. 
For $t{\bar t}$ production another class of corrections, subleading Coulomb terms, were shown to be very small 
in \cite{NKRV}. The contribution of hard-gluon radiation terms becomes smaller near threshold, where there 
is limited available energy. Each process needs to be studied
separately because of different kinematics, proximity to threshold, and color structures, 
and for each process the dominant terms need to be identified. For $s$-channel single top production, 
which is the process studied in this paper, the soft terms are dominant and they provide an excellent approximation 
to the complete cross section, which is why they are studied in detail here.

The NNLO expansion of the resummed cross section after inversion to 
momentum space is
\beqa
{\hat{\sigma}}^{(2)}&=&\sigma^B \frac{\alpha_s^2(\mu_R)}{\pi^2}
\left\{\frac{1}{2}c_3^2\, {\cal D}_3(s_4) + 
\left[\frac{3}{2}c_3 c_2-\frac{\beta_0}{4} c_3
+\frac{\beta_0}{8} A^{(1)}\right]  {\cal D}_2(s_4) \right.
\nonumber \\ && \hspace{-5mm} 
{}+\left[c_3 c_1+c_2^2-\zeta_2 c_3^2
-\frac{\beta_0}{2} T_2+\frac{\beta_0}{4} c_3 
\ln\left(\frac{\mu_R^2}{m_t^2}\right)+3 A^{(2)}
+\frac{\beta_0}{4} B^{(1)}
+4 \Gamma_{S\, 12}^{(1)} \Gamma_{S\, 21}^{(1)}\right] {\cal D}_1(s_4)
\nonumber \\ && \hspace{-5mm}  
{}+\left[c_2 c_1-\zeta_2 c_3 c_2+\zeta_3 c_3^2
+\frac{\beta_0}{4} c_2 \ln\left(\frac{\mu_R^2}{s}\right) \right. 
-\frac{\beta_0}{2} A^{(1)} \ln^2\left(\frac{m_t^2-t}{m_t^2}\right)
\nonumber \\ && \hspace{-5mm} \quad \quad  
{}-\frac{\beta_0}{2} A^{(1)} \ln^2\left(\frac{m_t^2-u}{m_t^2}\right)
+\left(-2 A^{(2)}+\frac{\beta_0}{2} D^{(1)}\right) 
\ln\left(\frac{(m_t^2-t)(m_t^2-u)}{m_t^4}\right)
\nonumber \\ && \hspace{-5mm}  \quad \quad
{}+B^{(2)}+3 D^{(2)}
+\frac{\beta_0}{4}  A^{(1)} \ln^2\left(\frac{\mu_F^2}{s}\right) 
-2 A^{(2)} \ln\left(\frac{\mu_F^2}{s}\right)
+\frac{3\beta_0}{8} A^{(1)} \ln^2\left(\frac{m_t^2}{s}\right)
\nonumber \\ && \hspace{-5mm}  \quad \quad \left. \left.
{}-\left(A^{(2)}+\frac{\beta_0}{4}
(B^{(1)}+2 D^{(1)})\right) \ln\left(\frac{m_t^2}{s}\right)
+2 \Gamma_{S\, 11}^{(2)}
+4 \Gamma_{S\, 12}^{(1)} \, \Gamma_{S\, 21}^{(1)} 
\ln\left(\frac{m_t^2}{s}\right)
\right]  {\cal D}_0(s_4) \right\}
\nonumber \\ 
\label{NNLOapprox}
\eeqa
where $\beta_0=(11 C_A-2 n_f)/3$ is the lowest-order beta function 
and all other quantities have been defined previously.
Note that all NNLO soft-gluon corrections are derived from the 
NNLL resummed cross section, i.e. the coefficients of all powers 
of logarithms in $s_4$ are given, from ${\cal D}_3(s_4)$ down 
to ${\cal D}_0(s_4)$. In Ref. \cite{NKst} where NLL accuracy 
was attained, only the coefficients of ${\cal D}_3(s_4)$ and ${\cal D}_2(s_4)$ 
were fully determined. 
Thus, at NNLL accuracy the theoretical improvement over NLL
is significant.
In the notation of Ref. \cite{NKRV}, 
where logarithmic accuracy in the expansion rather than the resummed 
exponent was used, the expansion from NNLL resummation is a NNLO-NNNLL result. 
To be clear, in the following 
sections we will use the notation NLL and NNLL to denote 
the corresponding accuracy in the 
resummed exponent, as we have done in this section.
As discussed in \cite{NKst,NNNLO} additional $\delta(s_4)$ terms 
involving $\zeta_2$ and $\zeta_3$ constants from the inversion to 
momentum space as well as $\delta(s_4)$ terms involving the factorization 
and renormalization scales are also computed.

In the following sections we add the NNLO soft-gluon terms of 
Eq. (\ref{NNLOapprox}) to the NLO cross section to derive an 
approximate NNLO cross section for $s$-channel 
single top and single antitop production at the Tevatron and LHC.

\mysection{Single top or antitop production at the Tevatron}

We begin our numerical study for $s$-channel single top quark production 
in proton-antiproton collisions at the Tevatron with $\sqrt{S}=1.96$ TeV. 
We note that the results for single antitop production at 
the Tevatron are identical. 
We use the MSTW2008 NNLO parton distribution functions (pdf) 
\cite{MSTW2008} in the calculation of the hadronic cross section. 

\begin{figure}
\begin{center}
\includegraphics[width=10cm]{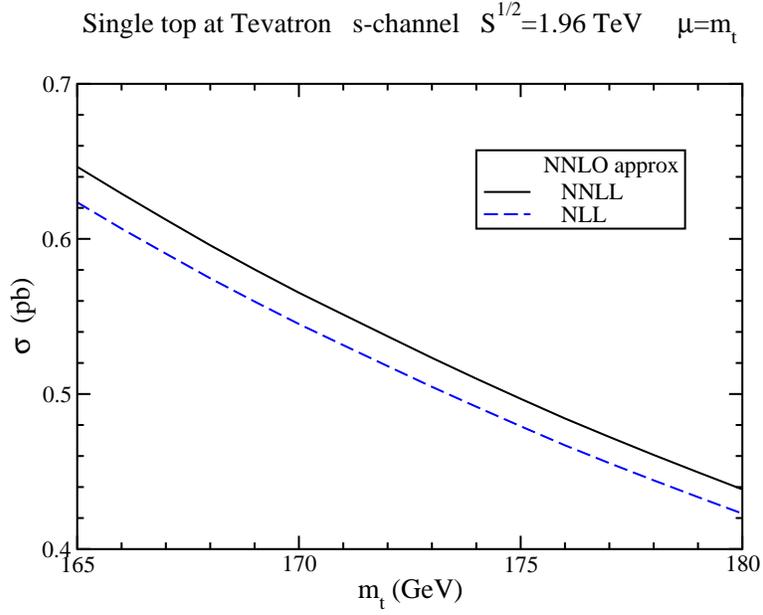}
\caption{The cross section for $s$-channel single top quark production 
at the Tevatron with $\sqrt{S}=1.96$ TeV and MSTW2008 NNLO pdf.}
\label{Tevcs}
\end{center}
\end{figure}

In Fig. \ref{Tevcs} we plot the NNLO approximate cross section for $s$-channel 
single top quark production at the Fermilab Tevatron as a function of top 
quark mass in the range 165 GeV $\le m_t \le$ 180 GeV. 
The factorization and renormalization scales are set equal to each other 
and this common scale, denoted by $\mu$, is set equal to the top quark mass.
Results are shown for the NNLO expansion from both NLL and NNLL resummation. 
The NLL result uses the expressions in \cite{NKst} while the NNLL 
result uses the new expression in Eq. (\ref{NNLOapprox}).
It is clear that the approximate NNLO cross section is larger at NNLL than 
at NLL, i.e. the additional NNLL numerical contributions are positive. 

\begin{figure}
\begin{center}
\includegraphics[width=10cm]{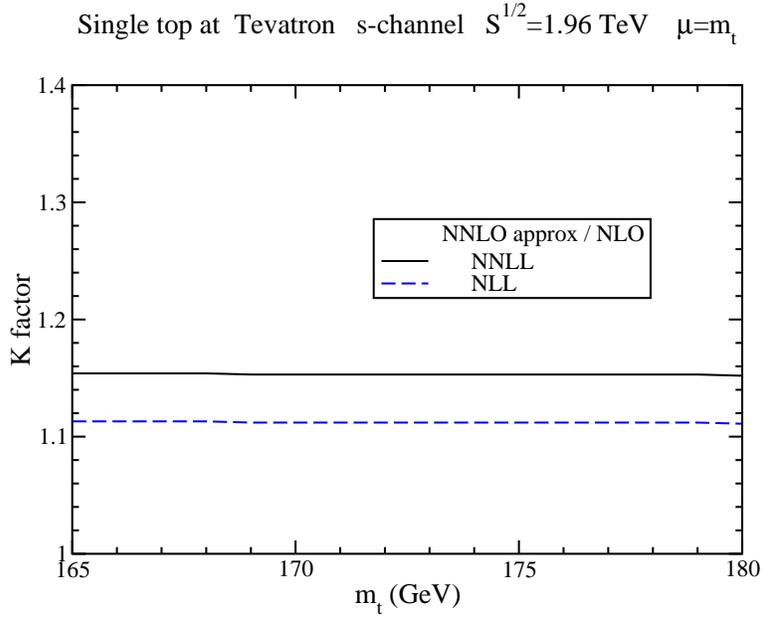}
\caption{The K factor for $s$-channel single top quark 
production at the Tevatron with $\sqrt{S}=1.96$ TeV.}
\label{TevK}
\end{center}
\end{figure}

It is important to know the additional contribution of the NNLO soft-gluon 
corrections at both NLL and NNLL accuracy relative to the NLO cross section. 
The corresponding $K$ factors, defined as 
the ratio of the NNLO approximate cross section to the NLO cross section, 
are displayed in Fig. \ref{TevK}. It is clear that the $K$ factors are quite 
insensitive to the value of the top quark mass. At NLL there is an 11\% 
enhancement over NLO, while at NNLL there is a 15\% enhancement over NLO. 
Thus the enhancement from soft-gluon corrections 
is quite significant at both NLL and NNLL accuracy.  
Also the new NNLL contributions increase the approximate NNLO cross section 
of NLL accuracy by an additional 3.7\%.

\begin{table}[htb]
\begin{center}
\begin{tabular}{|c|c|c|c|c|} \hline
\multicolumn{5}{|c|}{NNLO approx (NNLL) single top $s$-channel cross section (pb)} \\ \hline
$m_t$ (GeV) & Tevatron 1.96 TeV & LHC 7 TeV & LHC 10 TeV & LHC 14 TeV \\ \hline
170 & 0.565 & 3.39 & 5.50 & 8.45 \\ \hline 
171 & 0.551 & 3.31 & 5.38 & 8.27 \\ \hline 
172 & 0.537 & 3.24 & 5.27 & 8.10 \\ \hline 
173 & 0.523 & 3.17 & 5.16 & 7.93 \\ \hline 
174 & 0.510 & 3.10 & 5.05 & 7.76 \\ \hline 
175 & 0.497 & 3.03 & 4.94 & 7.60 \\ \hline 
\end{tabular}
\caption[]{The single top quark $s$-channel production cross section in pb 
in $p \overline p$ collisions at the Tevatron with $\sqrt{S}=1.96$ TeV, 
and in $pp$ collisions at the LHC with $\sqrt{S}=7$ TeV, 10 TeV, 
and 14 TeV, with 
$\mu=m_t$ and using the MSTW2008 NNLO pdf \cite{MSTW2008}.
The approximate NNLO results are shown at NNLL accuracy.}
\label{table1}
\end{center}
\end{table}

Table 1 lists the values of the approximate NNLO cross section 
at NNLL accuracy for top quark masses 
between 170 GeV and 175 GeV and with $\mu=m_t$.
There are theoretical uncertainties associated with these values that 
arise from the dependence on the scale $\mu$ as well as from pdf errors.
The scale uncertainty is most commonly estimated by varying the scale 
by a factor of two, i.e. between $m_t/2$ and $2m_t$. For the approximate 
NNLO cross section at NNLL at the Tevatron the scale uncertainty is 
+0.1\% $-$1.0\%, 
which is a significant improvement over NLO \cite{bwhl,HCSY} as well as over 
the NLL approximation \cite{NKst}.
The pdf uncertainty is calculated using the 40 different 
MSTW2008 NNLO eigensets as provided by MSTW at 90\% confidence level (C.L.)
\cite{MSTW2008} which provides a conservative estimate of pdf error.  
For $s$-channel single top quark production at the Tevatron 
this 90\% C.L. pdf uncertainty is +5.7\% $-$5.3\%. 
If instead one uses the 68\% C.L. NNLO eigensets provided by MSTW, 
the pdf uncertainty of the cross section becomes considerably smaller, 
+2.7\% $-$2.4\%, but it is still larger than the scale uncertainty.
Clearly at Tevatron energies the pdf uncertainty dominates 
the theoretical uncertainty in our approximate NNLO cross section at NNLL 
whether one uses the conservative 90\% C.L. 
or the 68\% C.L. pdf eigensets.

The best current value of the top quark mass is $173$ GeV \cite{topmass}. 
For this top quark mass we write the cross section 
and its associated uncertainties expilicitly as 
\beq
\sigma_{\rm s-ch}^{\rm top}(m_t=173\, {\rm GeV}, \, \sqrt{S}=1.96\, {\rm TeV})
=0.523^{+0.001}_{-0.005}{}^{+ 0.030}_{-0.028} \; {\rm pb}
\eeq
where the first uncertainty is from scale variation and the second is 
the pdf uncertainty at 90\% C.L.

\mysection{Single top quark production at the LHC}

We continue with numerical results for $s$-channel single top quark 
production at the LHC. We present results for three different energies: 
the design energy of 14 TeV, the planned starting energy of 7 TeV, 
and a possible intermediate run at 10 TeV. Again we use the MSTW2008 
NNLO pdf \cite{MSTW2008}. We note that at the LHC the single top and single 
antitop cross sections are different. In this section we focus on single top 
production, and we discuss single antitop production in the following section.

\begin{figure}
\begin{center}
\includegraphics[width=10cm]{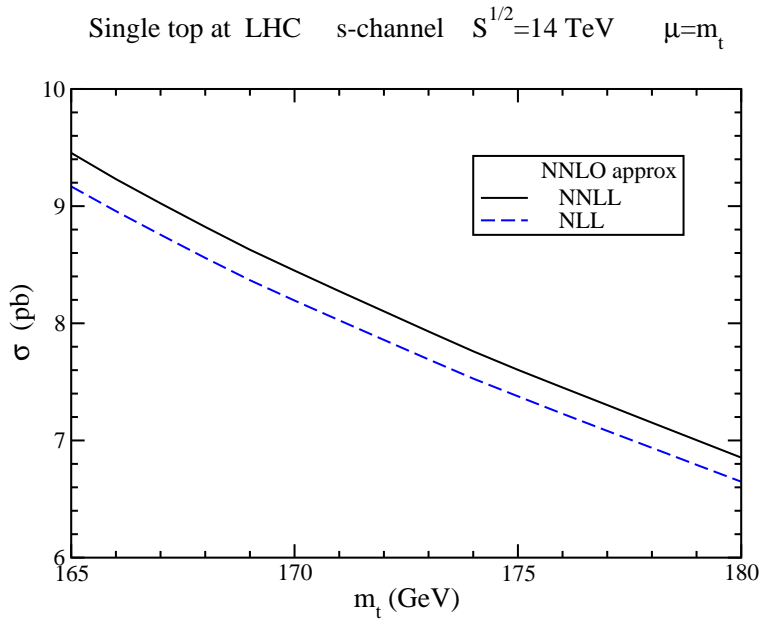}
\caption{The cross section for $s$-channel single top quark production 
at the LHC with $\sqrt{S}=14$ TeV and MSTW2008 NNLO pdf.}
\label{LHC14cs}
\end{center}
\end{figure}

In Fig. \ref{LHC14cs} we plot the NNLO approximate cross section 
for $s$-channel single top quark production at the LHC at its 
design energy of $\sqrt{S}=14$ TeV as a function of top quark mass. 
Results are shown for the NNLO expansion from both NLL and NNLL resummation. 
The NNLL result is larger than the NLL one.

\begin{figure}
\begin{center}
\includegraphics[width=10cm]{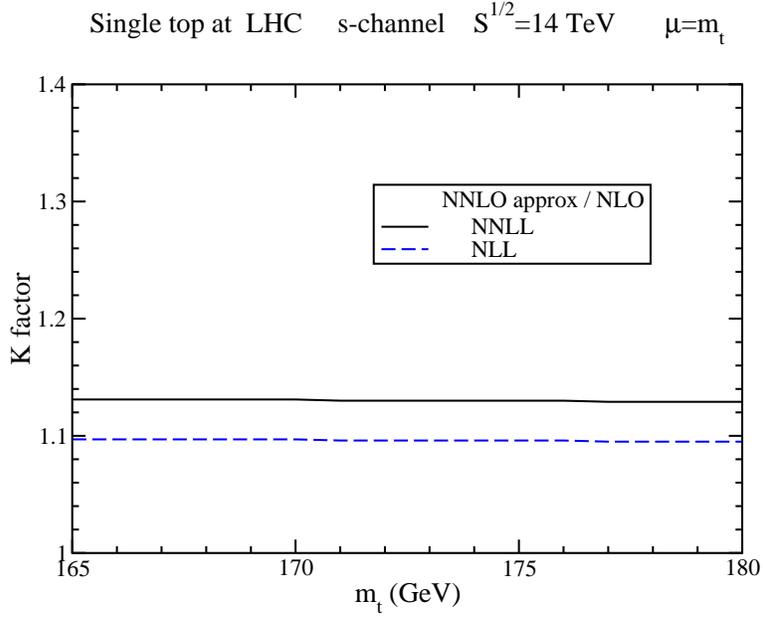}
\caption{The K factor for $s$-channel single top quark production at the LHC
with $\sqrt{S}=14$ TeV.}
\label{LHC14K}
\end{center}
\end{figure}

The $K$ factor, i.e. the ratio of the NNLO approximate cross section to
the NLO cross section, is displayed in Fig. \ref{LHC14K} at both NLL and NNLL.
Again the $K$ factors are quite insensitive to the value of the top quark mass.
At NLL there is nearly a 10\% enhancement over NLO, while at NNLL there is a
13\% enhancement over NLO. The enhancement from soft-gluon corrections
is similar to that for Tevatron collisions and is again quite
significant at both NLL and NNLL accuracy.

Table 1 lists the NNLO approximate cross section at NNLL accuracy 
for top quark masses between 170 GeV and 175 GeV 
for $\mu=m_t$ at 14 TeV.
The scale uncertainty of the results is $\pm 1.8$\%  and the  pdf uncertainty
at 90\% C.L. is +3.9\% $-$3.5\%, which is about twice 
as big as the scale uncertainty, while at 68\% C.L. it is +2.0\% $-$2.2\%. 
For a top quark mass of 173 GeV the explicit result is 
\beq
\sigma_{\rm s-ch}^{\rm top}(m_t=173\, {\rm GeV}, \, \sqrt{S}=14\, {\rm TeV})
=7.93 \pm 0.14 {}^{+ 0.31}_{-0.28} \; {\rm pb}
\eeq
where the first uncertainty is from scale variation and the second 
is from the pdf error at 90\% C.L. 

\begin{figure}
\begin{center}
\includegraphics[width=10cm]{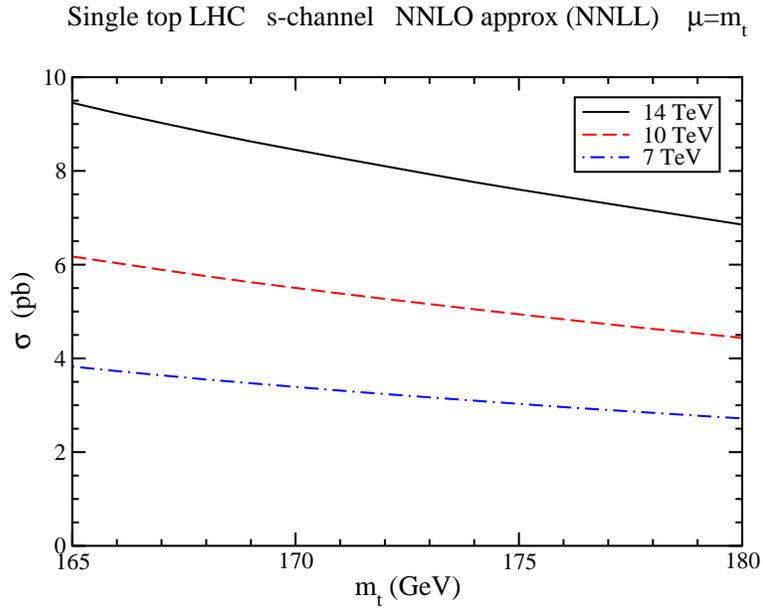}
\caption{The cross section for $s$-channel single top quark production 
at the LHC with $\sqrt{S}=7$ TeV, 10 TeV, and 14 TeV, and MSTW2008 NNLO pdf.}
\label{LHC7n10n14}
\end{center}
\end{figure}

Fig. \ref{LHC7n10n14} shows the NNLO approximate cross section at NNLL 
accuracy for $s$-channel single top quark
production at the LHC at the starting energy of $\sqrt{S}=7$ TeV and 
also at 10 TeV and at 14 TeV. 
The enhancement over NLO at 7 TeV and 10 TeV is very similar to that at 
14 TeV, over 13\%. 

Results for the cross section with $\mu=m_t$ at 10 TeV and 7 TeV 
are also displayed in Table 1.
At 10 TeV the scale uncertainty of the results is $\pm 1.8$\%  while  
the pdf uncertainty at 90\% C.L. is +3.9\% $-$2.8\% and at 68\% C.L. it 
is +2.2\% $-$1.4\%.
For a top mass of 173 GeV we have  
\beq
\sigma_{\rm s-ch}^{\rm top}(m_t=173\, {\rm GeV}, \, \sqrt{S}=10\, {\rm TeV})
=5.16 \pm 0.09 {}^{+ 0.20}_{-0.14} \; {\rm pb}
\eeq
where the first uncertainty is from scale variation and the second 
from the pdf at 90\% C.L.

At 7 TeV the scale uncertainty is $\pm 1.9$\%  while 
the pdf uncertainty is +4.2\% $-$3.1\% at 90\% C.L. and 
+2.2\% $-$1.6\% at 68\% C.L.
For $m_t=173$ GeV we have 
\beq
\sigma_{\rm s-ch}^{\rm top}(m_t=173\, {\rm GeV}, \, \sqrt{S}=7\, {\rm TeV})
=3.17 \pm 0.06 {}^{+ 0.13}_{-0.10} \; {\rm pb}
\eeq
where the first uncertainty is from scale variation and the second 
from the pdf at 90\% C.L.

\begin{figure}
\begin{center}
\includegraphics[width=10cm]{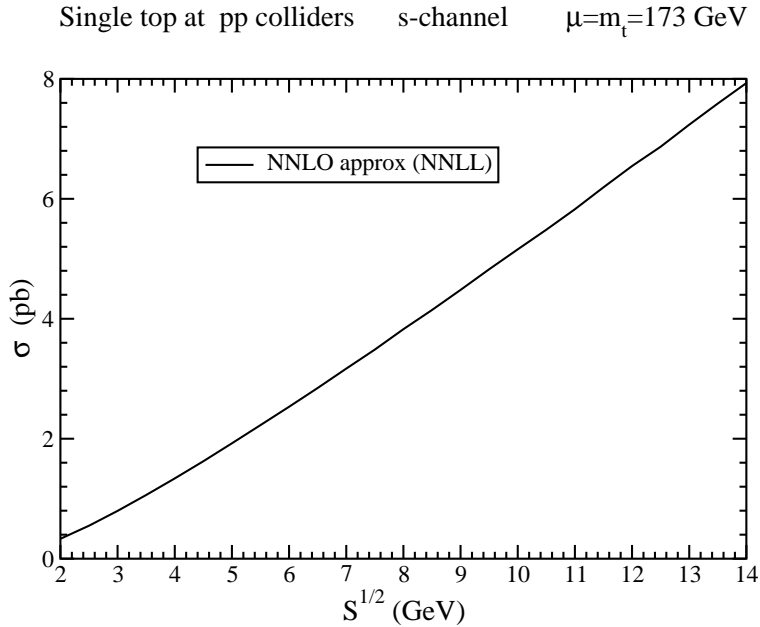}
\caption{The cross section for $s$-channel single top quark production 
at the LHC for energies $2 \le \sqrt{S} \le 14$ TeV.}
\label{LHCs}
\end{center}
\end{figure}

The dependence of the NNLO approximate cross section at NNLL accuracy 
on the LHC energy is plotted in Fig. \ref{LHCs} for the range  
$2 \le \sqrt{S} \le 14$ TeV with $m_t=173$ GeV. We see that 
the cross section at 14 TeV is about twenty four times bigger than at 2 TeV. 

\mysection{Single antitop production at the LHC}

We continue with results for single antitop production at the LHC 
in the $s$ channel. 

\begin{figure}
\begin{center}
\includegraphics[width=10cm]{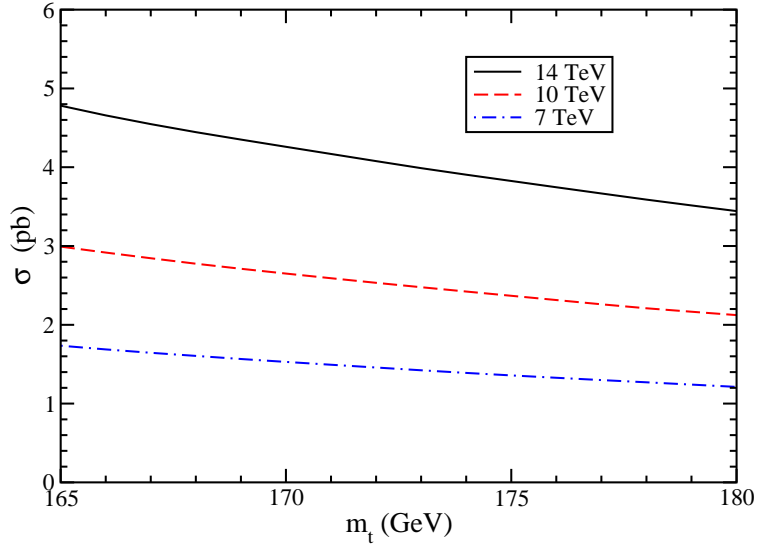}
\caption{The cross section for $s$-channel single antitop production at the LHC
with $\sqrt{S}=7$ TeV, 10 TeV, and 14 TeV, and MSTW2008 NNLO pdf.}
\label{LHCa7n10n14}
\end{center}
\end{figure}

Fig. \ref{LHCa7n10n14} shows the NNLO approximate cross section at NNLL 
accuracy for $s$-channel single antitop production at the LHC for 
energies of 7 TeV, 10 TeV, and 14 TeV, using the MSTW2008 NNLO pdf 
\cite{MSTW2008}. 
The cross sections are smaller than the corresponding ones for single top 
quark production by a factor of around two.

\begin{table}[htb]
\begin{center}
\begin{tabular}{|c|c|c|c|c|} \hline
\multicolumn{4}{|c|}{NNLO approx (NNLL) single antitop $s$-channel 
cross section (pb)} \\ \hline
$m_t$ (GeV) & \hspace{2mm} LHC 7 TeV \hspace{2mm}& \hspace{2mm} LHC 10 TeV \hspace{2mm} & \hspace{-2mm} LHC 14 TeV \hspace{-2mm} \\ \hline
170 & 1.53 & 2.65 & 4.26 \\ \hline 
171 & 1.49 & 2.59 & 4.17 \\ \hline 
172 & 1.46 & 2.53 & 4.08 \\ \hline 
173 & 1.42 & 2.48 & 3.99 \\ \hline 
174 & 1.39 & 2.42 & 3.91 \\ \hline 
175 & 1.36 & 2.37 & 3.83 \\ \hline 
\end{tabular}
\caption[]{The single antitop $s$-channel production cross section in
$pp$ collisions at the LHC with $\sqrt{S}=7$ TeV, 10 TeV, and 14 TeV, 
with $\mu=m_t$ and using the MSTW2008 NNLO pdf \cite{MSTW2008}.
The approximate NNLO results are shown at NNLL accuracy.}
\label{table2}
\end{center}
\end{table}

Table 2 lists the values of the single antitop approximate NNLO cross section 
at NNLL accuracy in the $s$-channel for antitop masses 
between 170 GeV and 175 GeV and $\mu=m_t$ for the three LHC energies.

At 14 TeV the scale uncertainty is $\pm 1.3$\%.
The pdf uncertainty is +3.4\% $-$5.2\% at 90\% C.L. and
+1.7\% $-$3.4\% at 68\% C.L.
For $m_t=173$ GeV, we find
\beq
\sigma_{\rm s-ch}^{\rm antitop}(m_t=173\, {\rm GeV}, \, \sqrt{S}=14\, 
{\rm TeV})=3.99 \pm 0.05 {}^{+ 0.14}_{-0.21} \; {\rm pb} 
\eeq
where the first uncertainty is from scale variation 
and the second from the pdf at 90\% C.L.
 
At 10 TeV the scale uncertainty is $\pm 0.9$\% while 
the pdf uncertainty is +3.5\% $-$5.3\% at 90\% C.L. and
+1.5\% $-$3.3\% at 68\% C.L.
For $m_t=173$ GeV, we find
\beq
\sigma_{\rm s-ch}^{\rm antitop}(m_t=173\, {\rm GeV}, \, \sqrt{S}=10\, 
{\rm TeV})=2.48 \pm 0.02 {}^{+ 0.09}_{-0.13} \; {\rm pb} 
\eeq
where the first uncertainty is from scale variation
and the second from the pdf at 90\% C.L.

At 7 TeV the scale uncertainty is $\pm 0.7$\%, and 
the pdf uncertainty is +4.2\% -5.0\% at 90\% C.L. and
+1.9\% $-$2.6\% at 68\% C.L.
For $m_t=173$ GeV, we find
\beq
\sigma_{\rm s-ch}^{\rm antitop}(m_t=173\, {\rm GeV}, \, \sqrt{S}=7\, {\rm TeV})
=1.42 \pm 0.01 {}^{+ 0.06}_{-0.07} \; {\rm pb} 
\eeq
where the first uncertainty is from scale variation
and the second from the pdf at 90\% C.L.

\begin{figure}
\begin{center}
\includegraphics[width=10cm]{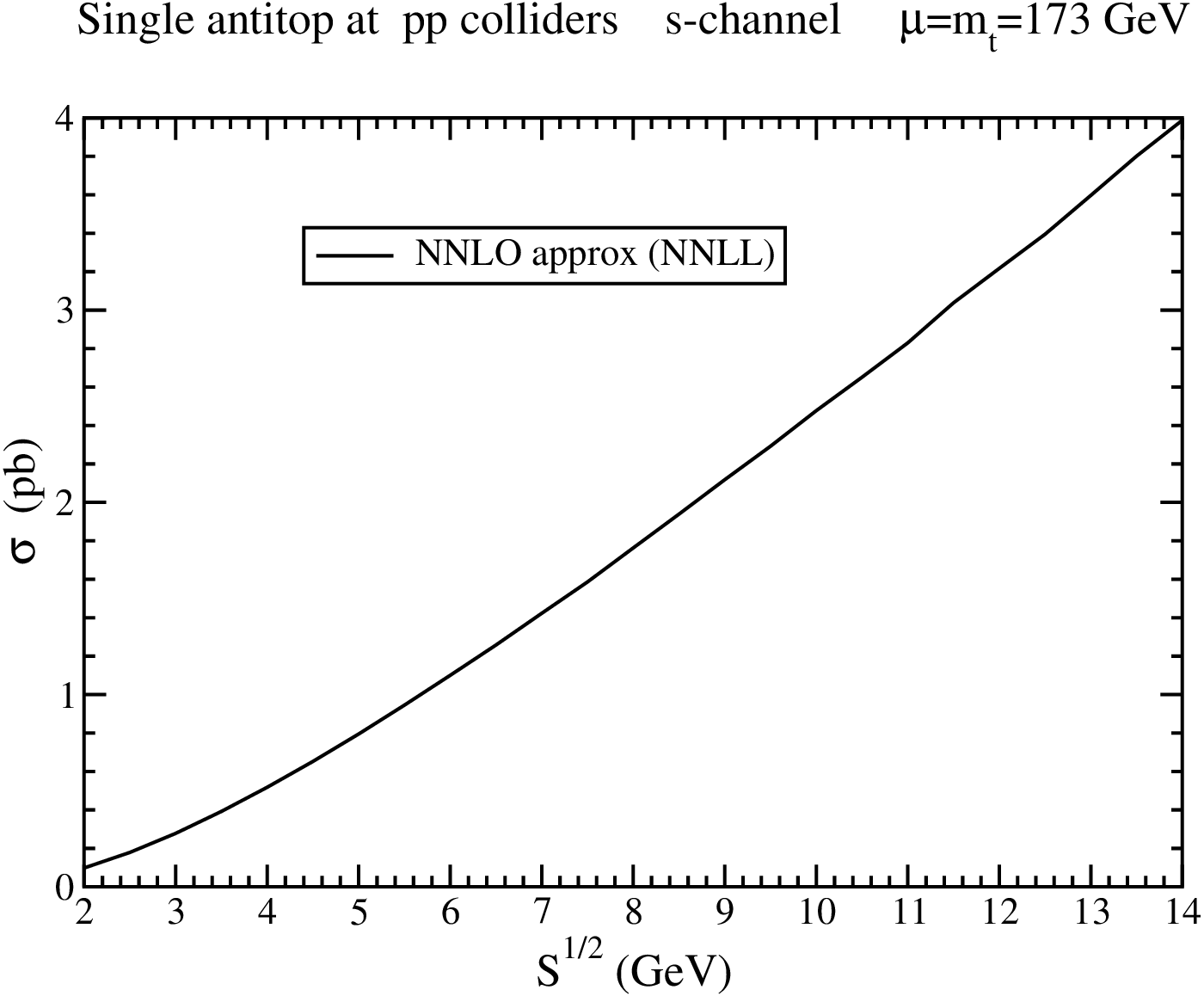}
\caption{The cross section for $s$-channel single antitop production at the LHC
for energies $2 \le \sqrt{S} \le 14$ TeV.}
\label{LHCas}
\end{center}
\end{figure}

The dependence of the NNLO approximate cross section at NNLL accuracy 
for single antitop production on the 
LHC energy is plotted in Fig. \ref{LHCas} for the range
$2 \le \sqrt{S} \le 14$ TeV and $m_t=173$ GeV. 

\mysection{Conclusion}
The single top quark production cross section in the $s$-channel 
receives significant contributions from soft-gluon corrections 
which increase the overall cross section and decrease the scale 
dependence of the theoretical prediction. The resummation of these 
corrections was performed at NNLL accuracy in this paper using an 
explicit calculation of the two-loop soft anomalous dimension. 
Approximate NNLO cross sections, which include NNLO soft-gluon 
corrections added to the NLO result, were calculated. Detailed 
numerical results were presented for single top and single antitop
production at the Tevatron and the LHC. The enhancement at the Tevatron 
over NLO is 15\% and at the LHC it is 13\%. 
In addition to the scale uncertainty, 
the pdf uncertainty was calculated using 90\% C.L. and 68\% C.L. eigensets. 
At 90\% C.L. the pdf uncertainty clearly dominates the theoretical error 
at both Tevatron and LHC energies.
The overall theoretical uncertainty of the approximate NNLO 
cross section from NNLL resummation is reduced compared to that at NLO
or at NLL accuracy.

\mysection*{Acknowledgements}
This work was supported by the National Science Foundation under 
Grant No. PHY 0855421.

\end{document}